\documentclass[letterpaper]{article}
\usepackage{iccc}
\usepackage{hyperref}

\usepackage[utf8]{inputenc} 
\usepackage[T1]{fontenc}    
\usepackage{url}            
\usepackage{verbatim}
\usepackage{times}
\usepackage{helvet}
\usepackage{courier}
\usepackage{paralist}
\usepackage{graphicx}
\usepackage{subcaption}
\usepackage[multiple]{footmisc}

\usepackage{xcolor}
\usepackage{soul}

\usepackage{tikz}
\usetikzlibrary{calc}

\makeatletter
\newif\if@anonymize

\@anonymizetrue    
\@anonymizefalse  

\if@anonymize
  \newcommand{\highlight@DoHighlight}{
    \fill [outer sep = -15pt, inner sep = 0pt, color=black]
          ($(begin highlight)+(0,8pt)$) rectangle ($(end highlight)+(0,-3pt)$) ;
  }

  \newcommand{\highlight@BeginHighlight}{
    \coordinate (begin highlight) at (0,0) ;
  }

  \newcommand{\highlight@EndHighlight}{
    \coordinate (end highlight) at (0,0) ;
  }

  \newdimen\highlight@previous
  \newdimen\highlight@current
  \newlength{\item@width}

  \DeclareRobustCommand*\anonymize{%
    \SOUL@setup
    \def\SOUL@preamble{%
      \begin{tikzpicture}[overlay, remember picture]
        \highlight@BeginHighlight
        \highlight@EndHighlight
      \end{tikzpicture}%
    }%
    \def\SOUL@postamble{%
      \begin{tikzpicture}[overlay, remember picture]
        \highlight@EndHighlight
        \highlight@DoHighlight
      \end{tikzpicture}%
    }%
    \def\SOUL@everyhyphen{%
      \discretionary{%
        \SOUL@setkern\SOUL@hyphkern
        \SOUL@sethyphenchar
        \tikz[overlay, remember picture] \highlight@EndHighlight ;%
      }{%
      }{%
        \SOUL@setkern\SOUL@charkern
      }%
    }%
    \def\SOUL@everyexhyphen##1{%
      \SOUL@setkern\SOUL@hyphkern
      \settowidth{\item@width}{##1}%
      \makebox[\item@width]{}%
      \discretionary{%
        \tikz[overlay, remember picture] \highlight@EndHighlight ;%
      }{%
      }{%
        \SOUL@setkern\SOUL@charkern
      }%
    }%
    \def\SOUL@everysyllable{%
      \begin{tikzpicture}[overlay, remember picture]
        \path let \p0 = (begin highlight), \p1 = (0,0) in \pgfextra
          \global\highlight@previous=\y0
          \global\highlight@current =\y1
        \endpgfextra (0,0) ;
        \ifdim\highlight@current < \highlight@previous
          \highlight@DoHighlight
          \highlight@BeginHighlight
        \fi
      \end{tikzpicture}%
      \settowidth{\item@width}{\the\SOUL@syllable}%
      \makebox[\item@width]{}%
      \tikz[overlay, remember picture] \highlight@EndHighlight ;%
    }%
    \SOUL@
  }

  \newcommand{\highlightwhite@DoHighlightwhite}{
    \fill [outer sep = -15pt, inner sep = 0pt, color=white]
          ($(begin highlightwhite)+(0,8pt)$) rectangle ($(end highlightwhite)+(0,-3pt)$) ;
  }

  \newcommand{\highlightwhite@BeginHighlightwhite}{
    \coordinate (begin highlightwhite) at (0,0) ;
  }

  \newcommand{\highlightwhite@EndHighlightwhite}{
    \coordinate (end highlightwhite) at (0,0) ;
  }

  \newdimen\highlightwhite@previous
  \newdimen\highlightwhite@current

  \DeclareRobustCommand*\anonymizewhite{%
    \SOUL@setup
    \def\SOUL@preamble{%
      \begin{tikzpicture}[overlay, remember picture]
        \highlightwhite@BeginHighlightwhite
        \highlightwhite@EndHighlightwhite
      \end{tikzpicture}%
    }%
    \def\SOUL@postamble{%
      \begin{tikzpicture}[overlay, remember picture]
        \highlightwhite@EndHighlightwhite
        \highlightwhite@DoHighlightwhite
      \end{tikzpicture}%
    }%
    \def\SOUL@everyhyphen{%
      \discretionary{%
        \SOUL@setkern\SOUL@hyphkern
        \SOUL@sethyphenchar
        \tikz[overlay, remember picture] \highlightwhite@EndHighlightwhite ;%
      }{%
      }{%
        \SOUL@setkern\SOUL@charkern
      }%
    }%
    \def\SOUL@everyexhyphen##1{%
      \SOUL@setkern\SOUL@hyphkern
      \settowidth{\item@width}{##1}%
      \makebox[\item@width]{}%
      \discretionary{%
        \tikz[overlay, remember picture] \highlightwhite@EndHighlightwhite ;%
      }{%
      }{%
        \SOUL@setkern\SOUL@charkern
      }%
    }%
    \def\SOUL@everysyllable{%
      \begin{tikzpicture}[overlay, remember picture]
        \path let \p0 = (begin highlightwhite), \p1 = (0,0) in \pgfextra
          \global\highlightwhite@previous=\y0
          \global\highlightwhite@current =\y1
        \endpgfextra (0,0) ;
        \ifdim\highlightwhite@current < \highlightwhite@previous
          \highlightwhite@DoHighlightwhite
          \highlightwhite@BeginHighlightwhite
        \fi
      \end{tikzpicture}%
      \settowidth{\item@width}{\the\SOUL@syllable}%
      \makebox[\item@width]{}%
      \tikz[overlay, remember picture] \highlightwhite@EndHighlightwhite ;%
    }%
    \SOUL@
  }
\else
  \newcommand{\anonymize}[1]{#1}
  \newcommand{\anonymizewhite}[1]{#1}
\fi
\makeatother

\newcommand\blfootnote[1]{%
  \begingroup
  \renewcommand\thefootnote{}\footnote{#1}%
  \addtocounter{footnote}{-1}%
  \endgroup
}

\title{NONOTO: A Model-agnostic Web Interface for Interactive Music Composition by Inpainting}

\author{
  \anonymizewhite{Th\'{e}is Bazin}  \\
  \anonymizewhite{Sony CSL}\\
  \anonymizewhite{Paris, France}\\
  \anonymizewhite{\texttt{theis.bazin@outlook.com}}
  \And
  \anonymizewhite{Ga\"{e}tan Hadjeres} \\
  \anonymizewhite{Sony CSL}\\
  \anonymizewhite{Paris, France}\\
  \anonymizewhite{\texttt{gaetan.hadjeres@sony.com}}
}

\begin{document}
\maketitle
\begin{NoHyper}
\blfootnote{Published as a conference paper at the 10th International Conference on Computational Creativity (ICCC 2019), UNC Charlotte, North Carolina.}
\end{NoHyper}

\begin{abstract}
\begin{quote}
Inpainting-based generative modeling allows for stimulating human-machine interactions by letting users perform stylistically coherent local editions to an object using a statistical model.
We present \emph{NONOTO}, a new interface for interactive music generation based on inpainting models. It is aimed both at researchers, by offering a simple and flexible API allowing them to connect their own models with the interface, and at musicians by providing industry-standard features such as audio playback, real-time MIDI output and straightforward synchronization with DAWs using Ableton Link.
\end{quote}
\end{abstract}


\subsubsection{Keywords}

interfaces, generative models, inpainting, interactive music generation, web technologies, open-source software

\section{Introduction}
We present a web-based interface that allows users to compose symbolic music in an interactive way using generative models for music. We strongly believe that such models only reveal their potential when actually used by artists and creators. While generative models for music have been around for a while~\cite{2012arXiv1206.6392B,hadjeres17a,roberts2018hierarchical}, the conception of A.I.-based interactive interfaces designed for music creators is still burgeoning. We contribute to this emerging area by providing a general web interface for many music generation models so that researchers in the domain can easily test and promote their works in actual music production and performance settings. This desire follows from the seminal work by \citeauthor{2015arXiv151101844T:theis:note_evaluation_generative_models}, in which the authors advocate that quantitative evaluation of generative models in an unambiguous way is hard and that "generative models need to be evaluated directly with respect to the application(s) they were intended for"~\cite{2015arXiv151101844T:theis:note_evaluation_generative_models}.
Lastly, we hope that the present work will contribute in making A.I.-assisted composition accessible to a wider audience, from non musicians to professional musicians, helping bridge the gap between these communities.

Drawing inspiration from recent advances in interactive interfaces for image restoration and editing~\cite{pix2pix2016,2019arXiv190206838J:sc_fegan,DBLP:journals/corr/abs-1806-03589:freeform_image_inpainting}, we focus on providing an interface for ``inpainting'' models for symbolic music, which are models that are able to recompose \emph{a portion of a score} given all the other portions. The reason is that such models
are more suited for an \emph{interactive} use (compared to models generating a full score all at once) and let users play an active part in the compositional process. As an outcome, users can feel that the
composition is the result of their work and not just something created by the machine. Furthermore, allowing quick exploration of musical ideas in a playful setting can enhance creativity and provide accessibility: the "technical part" of the composition is taken care of by the generative model which allows musicians as well as non experts in music to express themselves more freely.



\paragraph{Contributions:}The key elements of novelty are:
\begin{inparaenum}[(a)]
    \item easy-to-use and intuitive interface for users,
    \item easy-to-plug interface for researchers allowing them to explore the potential of their music generation algorithms,
    \item web-based and model-agnostic framework,
    \item integration of existing music inpainting algorithms,
    \item novel paradigms for A.I.-assisted music composition and live performance,
    \item integration in professional music production environments.
\end{inparaenum}

The code for the interface is distributed under a GNU GPL license and available, along with ready-to-use packaged standalone applications and video demonstrations of the interface, on our GitHub\footnote{\url{https://github.com/SonyCSLParis/NONOTO}}.

\subsection{Existing approaches}

The proposed system is akin to the FlowComposer system~\cite{papadopoulos16:flow_composer} which offers to generate sheets of music by performing local updates (using Markov Models in their case). However, this interface does not exhibit the same level of interactivity as ours since no real-time audio nor MIDI playback is available, which limits the tool to solely studio usage and makes for a less spontaneous and reactive user experience.

The recent tools proposed by the Google Magenta team as part of their Magenta Studio effort~\cite{magenta_studio} are more aligned with our aims in this project: they offer a selection of Ableton Live plugins (using Max for Live) that make use of various symbolic music generation models for rhythm as well as melody~\cite{roberts2018hierarchical,DBLP:journals/corr/abs-1809-04281}. Similarly, the StyleMachine, developed by Metacreative Technologies~\footnote{\url{https://metacreativetech.com/products/stylemachine-lite/}}, is a proprietary tool that allows to generate midi-tracks into Ableton Live in various dance music styles using a statistical model trained on different stylistic corpora of electronic music. Yet these tools differ from ours in the generation paradigm used: they offer either continuation-based (the model generates the end of a sequence given the beginning) or complete generation (the model generates a full sequence, possibly given a template), thus breaking the flow of music on new generations. We believe that this limits their level of interactivity as opposed to local, inpainting-based models as ours, as mentioned previously. In particular, it hinders their usage in live, performance contexts.


\section{Suggested mode of usage}
\begin{figure}[t]
  \centering
  \subcaptionbox{\label{fig:1}}{\includegraphics[width=2.5in]{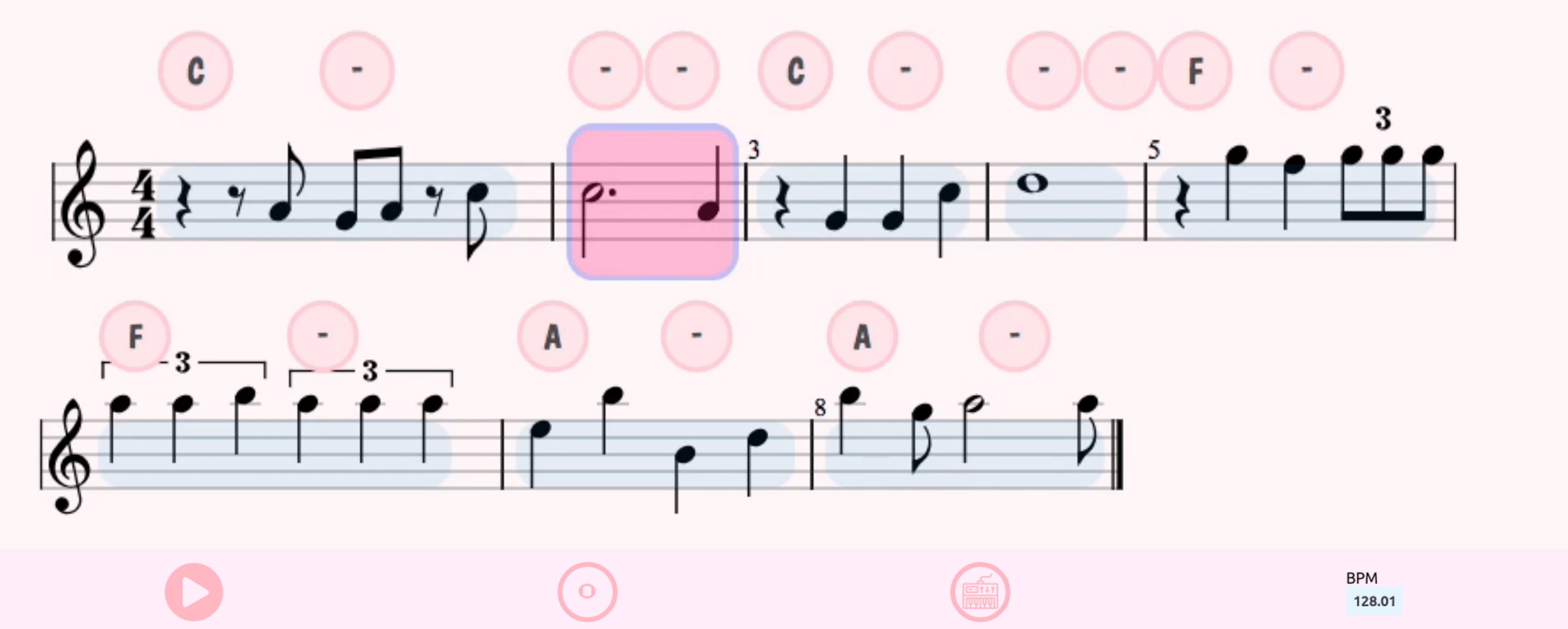}}\hfill%
  \subcaptionbox{\label{fig:2}}{\includegraphics[trim={0 0 0 200},clip, width=2.5in]{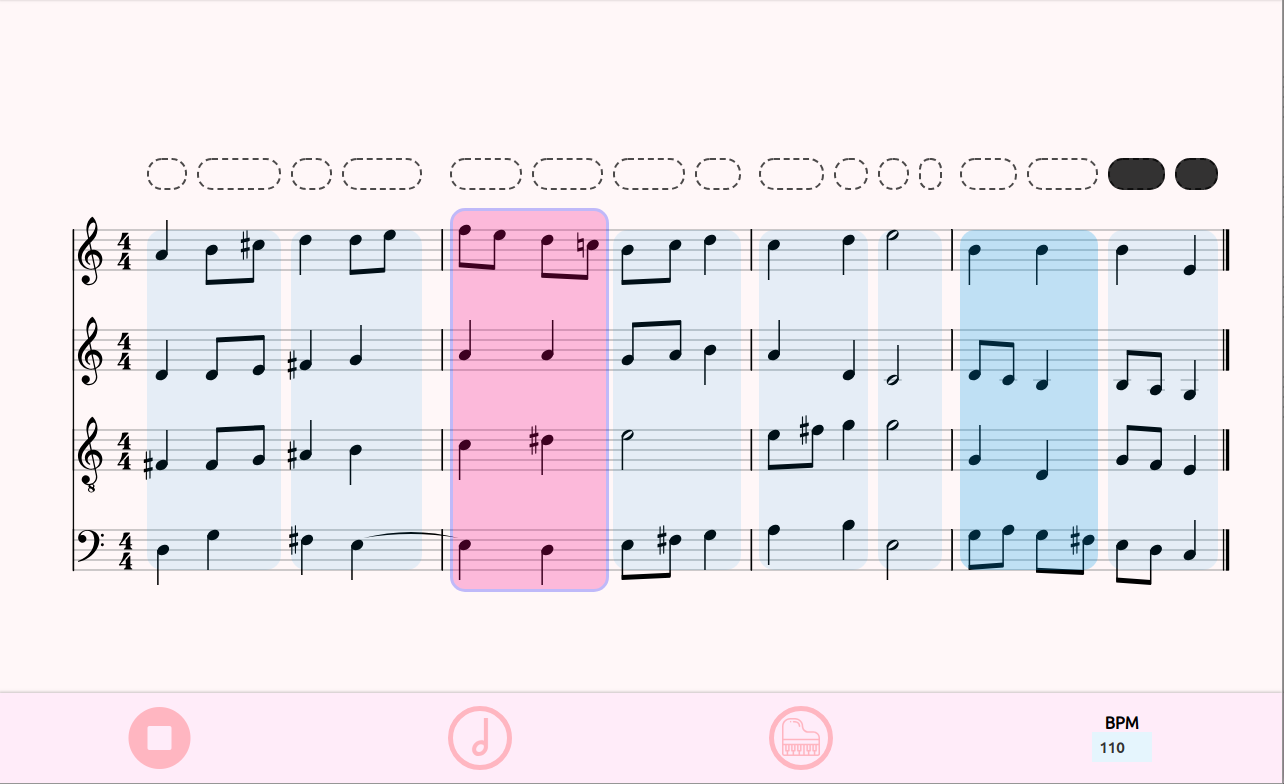}}\hfill%
  \caption{Our web interface used on different datasets: \ref{fig:1} melody and symbolic chords format, \ref{fig:2} four-part chorale music.}
  \label{fig:interface}
\end{figure}

The interface displays a musical score which loops forever as shown in Figure~\ref{fig:interface}. Users can then modify the score by regenerating any region only by clicking/touching it. The displayed musical score is updated instantly without interrupting the music playback. Other means of control are available depending on the specificity of the training dataset: we implemented, for instance, the positioning of fermatas in the context of Bach chorales generation or the control of the chord progression when generating Jazz leadsheets or Folk songs. These \emph{metadata} are sent, along with the sheet, to the generative models when performing re-generations.
The scores can be seamlessly integrated in a DAW 
so that the user (or even other users) can shape the sounds, add effects, play the drums or create live mixes. This creates a new jam-like experience in which the user controlling the A.I. can be seen as just one of the multiple instrument players. This interface thus has the potential to create a new environment for collaborative music production and performance.

Since our approach is flexible, our tool can be used in conjunction with other A.I.-assisted musical tools like Magenta Studio~\cite{magenta_studio} or the StyleMachine (from Metacreative Technologies). 

\section{Technology}
\subsection{Framework}

Our framework relies on two elements: an interactive web interface and a music inpainting server. This decoupling is strict so that researchers can easily plug-in new inpainting models with little overhead: it suffices to implement how the music inpainting model should function given a parti\-cular user input.
We make heavy use of modern web browser technologies, making for a modular and hackable code-base for artists and researchers, allowing e.g. to edit the interface to allow for some particular means of interaction or to add support for some new metadata specific to a given corpus.

\subsubsection{Interface}
The interface is based on OpenSheetMusicDisplay~\footnote{\url{https://github.com/opensheetmusicdisplay/opensheetmusicdisplay/}}, a TypeScript library aimed at rendering MusicXML sheets with vector graphics. Using Tone.js~\cite{mann15:interactive_music_tonejs}, a JavaScript library for real-time audio synthesis, we augmented OSMD with real-time audio playback capacities, allowing users to preview the generated music in real-time from within the interface. Furthermore, the audio playback is uninterrupted by re-generations, enabling a truly interactive experience.

\subsubsection{Generation back-end and communication}

For better interoperability, we rely on the MusicXML standard to communicate scores between the interface and the server.
The HTTP-based communication API then just consists in two commands that are required server-side:
\begin{itemize}
    \item A \texttt{/generate} command which expects the generation model to return a fresh new sheet of music in the MusicXML format to initialize a session,
    \item A \texttt{/timerange-change} command which takes as parameter the position of the interval to re-generate.
    The server is then expected to return an updated sheet with the chosen portion regenerated by the model using the current musical context.
\end{itemize}


\subsubsection{DAW integration}
In order for NONOTO to be readily usable in traditional music production and performance contexts, we implemented the possibility of integrating the generated scores in any DAW in real time.
To this end, we provide the user with the option of either rendering the generated sheet to audio in real-time from within the web interface using Tone.js or of routing it via MIDI to any virtual MIDI port on the host machine, using the JavaScript bindings to the Web MIDI API, WebMidi.js~\footnote{\url{https://github.com/djipco/webmidi}}. We also integrated support for Ableton Link~\footnote{\url{https://github.com/Ableton/link/}}\footnote{ \url{https://github.com/2bbb/node-abletonlink}}, an open-source technology developped by Ableton for easy synchronization of musical hosts on a local network, allowing to syncronize the inferface with e.g. Ableton Live.
Adding support for these technologies does not represent novel advances on our side \emph{per se}, yet, paired with the support of arbitrary generation back-ends, they allow to quickly test new generation models in a standard music production environment with minimal overwork and make for a beneficial tool for researchers -- and the first of its kind to our knowledge.

\section{Conclusion}

We have introduced NONOTO, an interactive, open-source and hackable interface for music generation using inpainting models. We invite researchers and artists alike to make it their own by developing new models or means of interacting with those. This high level of hackibility is to a large extent permitted by the wide range of technologies now offered in a very convenient fashion by modern web browsers, from which we draw heavily. Ultimately, we hope that providing tools such as ours with a strong focus on usability, accessibility, affordance and hackability will help shift the general perspective on machine learning for music creation, transitioning from the current and somewhat negative view of "robot music", replacing musicians, towards a more realistic and humbler view of it as A.I.-\emph{assisted} music.

\section{Ackowledgements}

The authors would like to thank the anonymous reviewers for their helpful comments.

\bibliographystyle{hapalike}
\bibliography{bibfile}

\end{document}